\documentclass{article}
\usepackage{amsmath,graphicx,mlspconf,hyperref}
\usepackage{amsfonts}
\usepackage{tikz}
\usetikzlibrary{shapes,arrows,positioning,calc}
\usepackage{url}

\copyrightnotice{978-1-5090-6341-3/17/\$31.00 {\copyright}2017 IEEE}

\toappear{2017 IEEE International Workshop on Machine Learning for Signal Processing, Sept.\ 25--28, 2017, Tokyo, Japan}

\title{Neural Network Alternatives to Convolutive Audio Models for Source Separation}
\twoauthors
  {Shrikant Venkataramani$^{\dagger}$, Cem Subakan$^{\dagger}$}
    {University of Illinois at Urbana Champaign \\
     \{svnktrm2,subakan2\}@illinois.edu \\
     $^{\dagger}$Authors with equal contribution
     }
  {Paris Smaragdis\sthanks{This work was supported by NSF grant 1453104.}}
    {University of Illinois at Urbana-Champaign \\ 
     Adobe Research \\
     paris@illinois.edu
    }

\tikzset{
block/.style = {draw, fill=white, rectangle, minimum height=3em, minimum width=3em},
tmp/.style  = {coordinate}, 
sum/.style= {draw, fill=white, circle, node distance=1cm},
input/.style = {coordinate},
output/.style= {coordinate},
pinstyle/.style = {pin edge={to-,thin,black}}
}

\begin{document}
\maketitle

\begin{abstract}
Convolutive Non-Negative Matrix Factorization model factorizes a given audio spectrogram using frequency templates with a temporal dimension. In this paper, we present a convolutional auto-encoder model that acts as a neural network alternative to convolutive NMF. Using the modeling flexibility granted by neural networks, we also explore the idea of using a Recurrent Neural Network in the encoder.  Experimental results on speech mixtures from TIMIT dataset indicate that the convolutive architecture provides a significant improvement in separation performance in terms of BSSeval metrics. 
\end{abstract}


\section{Introduction}
\label{sec:intro}

Non-negative matrix factorization~(NMF) of magnitude-spectrograms has been a very popular method of modeling sources for supervised source separation applications~\cite{smaragdis2014static, virtanen2015compositional}. NMF factorizes a matrix of non-negative elements~$\mathbf{X}\in \mathbb{R}_{M\times N}^{\geq0}$ as a product of the basis matrix~$\mathbf{W}\in \mathbb{R}_{M \times K}^{\geq{0}}$ and the activation matrix~$\mathbf{H}\in \mathbb{R}_{K \times N}^{\geq{0}}$. The notation $\mathbb{R}_{M \times N}^{\geq0}$ represents the set of matrices of non-negative elements of size $M \times N$, and $K$ represents the rank of the decomposition. In the case of audio signals, NMF is applied on audio spectrograms, where the columns of $\mathbf{W}$ act as representative basis vectors for the source. The rows of $\mathbf{H}$ indicate the activity of these basis vectors in time.

As shown in~~\cite{smaragdis2017aneural}, the notion of non-negative audio modeling can be easily generalized by interpreting NMF as a neural network. We can interpret NMF as a non-negative auto-encoder in the following manner,
\begin{align}
    \text{$1^{\text{st}}$ layer:~ (Encoder)~}\mathbf{H} &= g(\mathbf{W^{\ddagger}} \cdot \mathbf{X}) \nonumber \\
    \text{$2^{\text{nd}}$ layer:~ (Decoder)~}\mathbf{X} &= g(\mathbf{W} \cdot \mathbf{H})
    \label{eq:nmfae}
\end{align}
Here, $\mathbf{X}$ represents the input spectrogram, $\mathbf{W^{\ddagger}}$ represents a form of pseudo-inverse of $\mathbf{W}$ and $g(.):\mathbf{R}\rightarrow \mathbf{R}^{\geq0}$ is an element-wise function that maps a real number to the space of positive real numbers. As before, the columns of $\mathbf{W}$ act as representative basis vectors and the corresponding rows of $\mathbf{H}$ indicate their respective activations. Although non-negativity of the network-parameters (models) is not explicitly guaranteed in this formulation, applying a suitable sparsity constraint allows the network to learn suitable non-negative models~\cite{smaragdis2017aneural}. Additionally, this interpretation enables a pathway to propose variants to this basic autoencoder stucture by exploiting the wealth of available neural net architectures that could potentially lead to superior separation performance.

Spectrograms of speech and audio signals incorporate temporal dependencies that span multiple time frames. However, NMF and its neural network equivalent are unable to explicitly utilize these cross-frame patterns available in a spectrogram. To alleviate this drawback, Smaragdis~\cite{smaragdis2007convolutive} proposed a convolutive version to NMF~(conv-NMF) that allows spectro-temporal patterns as representative basis elements. In this paper, we develop a neural network alternative to such convolutive audio models for supervised source separation. In doing so, we solve two fundamental issues associated with this task. (i) We develop a suitable neural network architecture to learn convolutive audio models in an adaptive manner. (ii) We then utilize the learned models to separate a source from a given mixture.

Several neural network architectures have been recently proposed for supervised source separation~\cite{grais2017single, venkataramani2017end, chandna2017monoaural}, where the networks are discriminatively trained. In other words, these networks operate directly on the mixtures and separate them into individual sources. Although discriminative training of a source separation network results in good seperation performance, the network is restricted to work on a particular type of mixture. The convolutional architecture that we mainly discuss in this paper is generatively trained on the magnitude spectrograms of clean utterances. Therefore, these networks are not restricted to the types of mixtures used in training. By the virtue of flexibility of neural networks, we also propose a variant where recurrent neural network is used.

The remainder of the paper is organized as follows. In section~\ref{sec:conv-nmf}, we develop an auto-encoder that can act as an equivalent to conv-NMF audio models. Section~\ref{sec:rnncnn} discusses a novel extension to the convolutional auto-encoder based on a cascade of recurrent and convolutional layers. Section~\ref{sec:ss} proposes a novel approach to utilize these models for supervised source separation. We evaluate these models in terms of their separation performance in section~\ref{sec:experiments} and conclude in section~\ref{sec:conclusion}.

\begin{figure}[t]
\centering
  \includegraphics[clip, trim = 0cm 0cm 0cm 0cm, width=\linewidth]{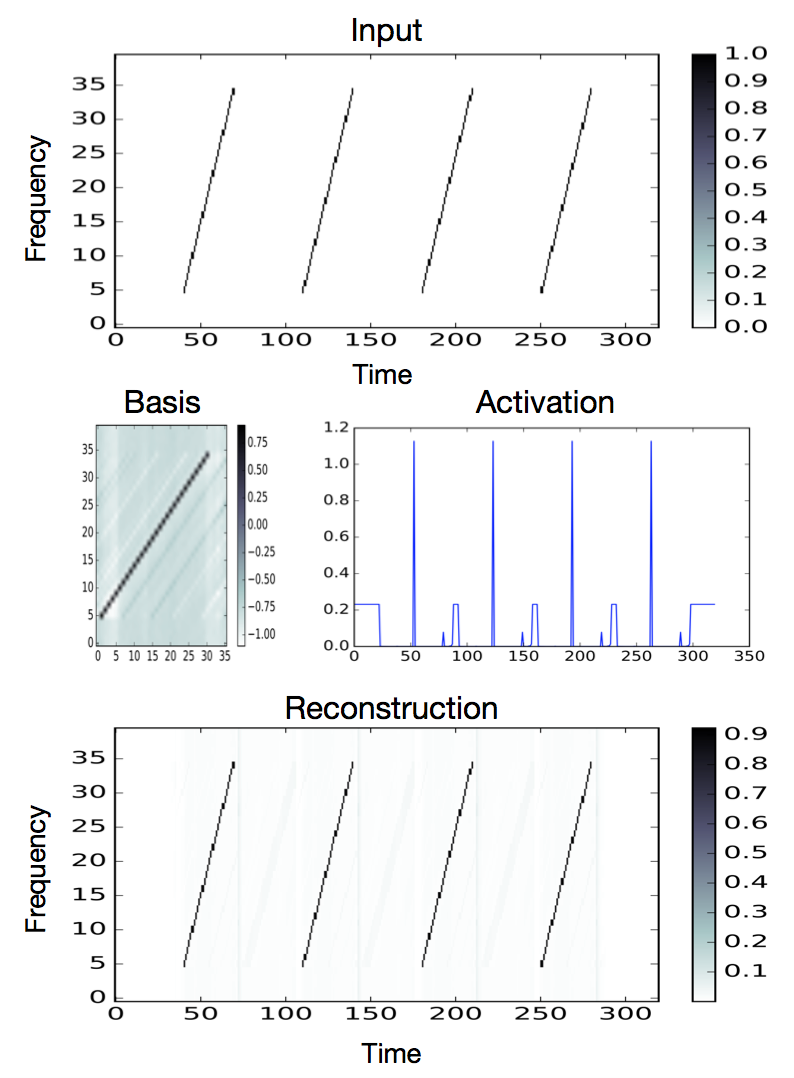}
  \caption{Basis decomposition of a toy-illustration obtained using a CNN-CNN auto-encoder.}~\label{fig:cnn_demo}
\end{figure}

\section{Non-negative Convolutional Auto-encoders}
\label{sec:conv-nmf}
\subsection{Network Architecture}
The convolutive NMF model~\cite{smaragdis2007convolutive} approximates a non-negative matrix~$\mathbf{X}\in \mathbb{R}_{M \times N}^{\geq0}$ as,
\begin{equation}
    \mathbf{X}(f,t) \approx \sum_{i=1}^{K} \sum_{k=0}^{T-1} \mathbf{W}_{i}(k,f)\cdot\mathbf{H}(i,t-k)
\end{equation}
Here, $\mathbf{W}_{i} \in \mathbb{R}_{M \times T}^{\geq0}$ acts as the $i^{\text{th}}$ spectro-temporal basis matrix out of $K$ such matrices and $\mathbf{H} \in \mathbb{R}_{K \times N}^{\geq0}$ contains the corresponding weights. The notation $\mathbf{X}(i,j)$ represents the element of $\mathbf{X}$ indexed by the $i^{\text{th}}$ row and the $j^{\text{th}}$ column. We can interpret this operation as a two-layer convolutional auto-encoder as follows,

\begin{align}
  \text{$1^{st}$ layer:~}\mathbf{H}(i,t) &= \sum_{j=0}^{M-1} \sum_{k=0}^{T-1} \mathbf{W}_{i}^{\ddagger}(j,k)\mathbf{X}(j,t-k) \nonumber \\
  \text{$2^{nd}$ layer:~}\hat{\mathbf{X}}(f,t) &= \sum_{i=1}^{K} \sum_{k=0}^{T-1} \mathbf{W}_{i}(k,f)\cdot\mathbf{H}(i,t-k)
    \label{eq:convnmfasnn}
\end{align}
subject to non-negativity of $\mathbf{W}_{i}$ and $\mathbf{H}$. Here, we assume that the convolutional layer filters~$\mathbf{W},~\mathbf{W}^{\ddagger}$ have a size of $M\times T$ where, $T$ represents the depth of the convolution and $M$ denotes the height of the input matrix~$\mathbf{X}$. In this representation, $\mathbf{W}_{i}$ and $\mathbf{H}$ correspond to the $i^{\text{th}}$ basis matrix and the activation matrix respectively. The filters of the first convolutional neural network~(CNN) act as inverse filters in defining the auto-encoder. In the remainder of this section, we will refer to the first convolutional layer as the ``encoder'' that estimates a code from the input representation. The second CNN layer generates an approximation of the input from the code and will be referred to as the ``decoder''. Finally, we will refer to this auto-encoder as the CNN-CNN auto-encoder~(CCAE). We can satisfy the non-negativity constraints by incorporating a non-linearity into the definitions of the encoder and the decoder. Thus,

\begin{align}
    \text{$1^{\text{st}}$ layer:~}\mathbf{H}(i,t) &= g\left(\sum_{j=0}^{M-1} \sum_{k=0}^{T-1} \mathbf{W}_{i}^{\ddagger}(j,k)\mathbf{X}(j,t-k)\right) \nonumber \\
    \text{$2^{\text{nd}}$ layer:~}\hat{\mathbf{X}}(f,t) &= g\left(\sum_{i=1}^{K} \sum_{k=0}^{T-1} \mathbf{W}_{i}(k,f)\cdot\mathbf{H}(i,t-k)\right)
    \label{eq:nmfcnncnn}
\end{align}

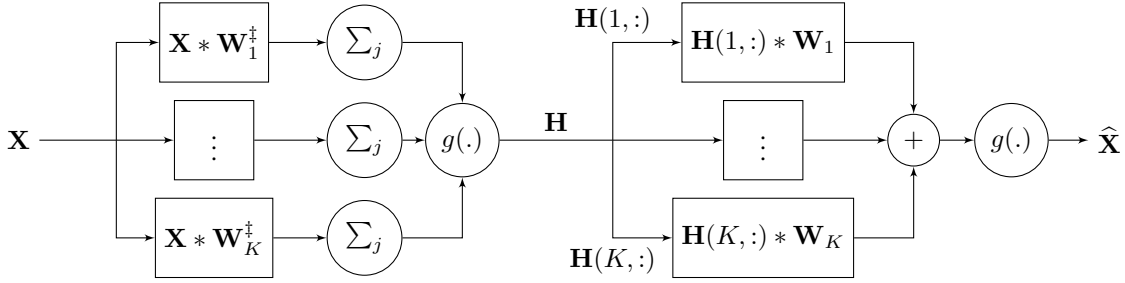
\begin{figure*}[t]
  \centering
  \begin{tikzpicture}[auto, node distance=2cm,>=latex']
      
      \node [,name=input] (input) {$\mathbf X$};
      \coordinate [, name=rinput, right of=input, node distance=1.3cm ](rinput) {};
      \node [block, right of=rinput, node distance=1.3cm] (controller) {$\vdots$};
      \node [sum, right of=controller, node distance=2cm] (controllers){$\sum_j$};
      \node [block, above of=controller,node distance=1.3cm] (up){$\mathbf X *\mathbf W^\ddagger_1$};
      \node [sum, right of=up, node distance=2cm] (ups) {$\sum_j$};
      \node [block, below of=controller,node distance=1.3cm] (rate) {$\mathbf X*\mathbf W^\ddagger_K$};
      \node [sum, right of=rate, node distance=2cm] (rates) {$\sum_j$};	

      \node [sum, right of=controllers,node distance=1.3cm] (sum2) {$g(.)$};
      \coordinate[, right of=sum2,node distance=2cm] (system) {};
      \node [block, right of=system, node distance=2cm] (dec2) {$\vdots$};
      \node [block, above of=dec2, node distance=1.3cm] (dec1) {$\mathbf H(1, :) * \mathbf W_1$};
      \node [block, below of=dec2, node distance=1.3cm] (dec3) {$\mathbf H(K, :) *\mathbf W_K$};
      \node [sum, right of=dec2, node distance=2cm] (sum3) {$+$};
      \node [sum, right of=sum3, node distance=1.3cm] (g2) {$g(.)$};
      \node [,right of=g2, node distance=1.3cm] (outp) {$\widehat{\mathbf X}$};

      \draw [-] (input) -- (rinput);
      \draw [->] (rinput) -- (controller); 
      \draw [->] (controller) -- (controllers);
      \draw [->] (controllers) -- (sum2);
      \draw [-] (sum2) --node{$\mathbf H$} (system);
      \draw [->] (rinput) |- (rate);
      \draw [->] (rate) -- (rates);
      \draw [->] (rates) -| (sum2);
      \draw [->] (rinput) |- (up);
      \draw [->] (up) -- (ups);
      \draw [->] (ups) -| (sum2);
      \draw [->] (system) -- node {} (dec2);
      \draw [->] (system) |- node {$\mathbf H(1,:)$} (dec1);
      \draw [->] (system) |- node[anchor=north]{$\mathbf H(K,:)$} (dec3);
      \draw [->] (dec1) -| (sum3);
      \draw [->] (dec2) -- (sum3);
      \draw [->] (dec3) -| (sum3);
      \draw [->] (sum3) -- (g2);
      \draw [->] (g2) -- (outp);
    \end{tikzpicture}
  \caption{Block Diagram of CNN-CNN Autoencoder} 
  \label{diag:cnncnn}
\end{figure*}

Here, the $g(.):\mathbb{R}\rightarrow \mathbb{R}^{\geq0}$ applies an element-wise non-linearity and ensures that the activation matrix and the reconstruction are non-negative. The block diagram of the whole CCAE is given in Figure \ref{diag:cnncnn}. In our experiments, we used the soft-plus function which is given by the formula $g(x) = \log(1+\exp(x))$ as the non-linearity. Using (\ref{eq:nmfcnncnn}), we now note some key points about the CCAE. (i) The output of the encoder gives the latent representation (analog of activations in NMF) of the decomposition. (ii) The filters of the decoder act as the spectro-temporal bases of the decomposition. (iii) We do not explicitly apply non-negativity constraints on the bases (decoder filters). Thus, the basis matrices can assume negative values. To train the auto-encoder, we minimize the KL-divergence between the input spectrogram~$\mathbf{X}$ and its reconstruction~$\hat{\mathbf{X}}$ given by,
\begin{equation}
    D\left(\mathbf{X},\hat{\mathbf{X}}\right) = \sum_{i,j} \mathbf{X}(i,j)\cdot \text{log}\frac{\mathbf{X}(i,j)}{\hat{\mathbf{X}}(i,j)} - \mathbf{X}(i,j) + \hat{\mathbf{X}}(i,j)
\end{equation}

\begin{figure}[ht!]
\centering
  \includegraphics[clip, trim = 0cm 0cm 0cm 0cm, width=\columnwidth, height = 0.8\columnwidth]{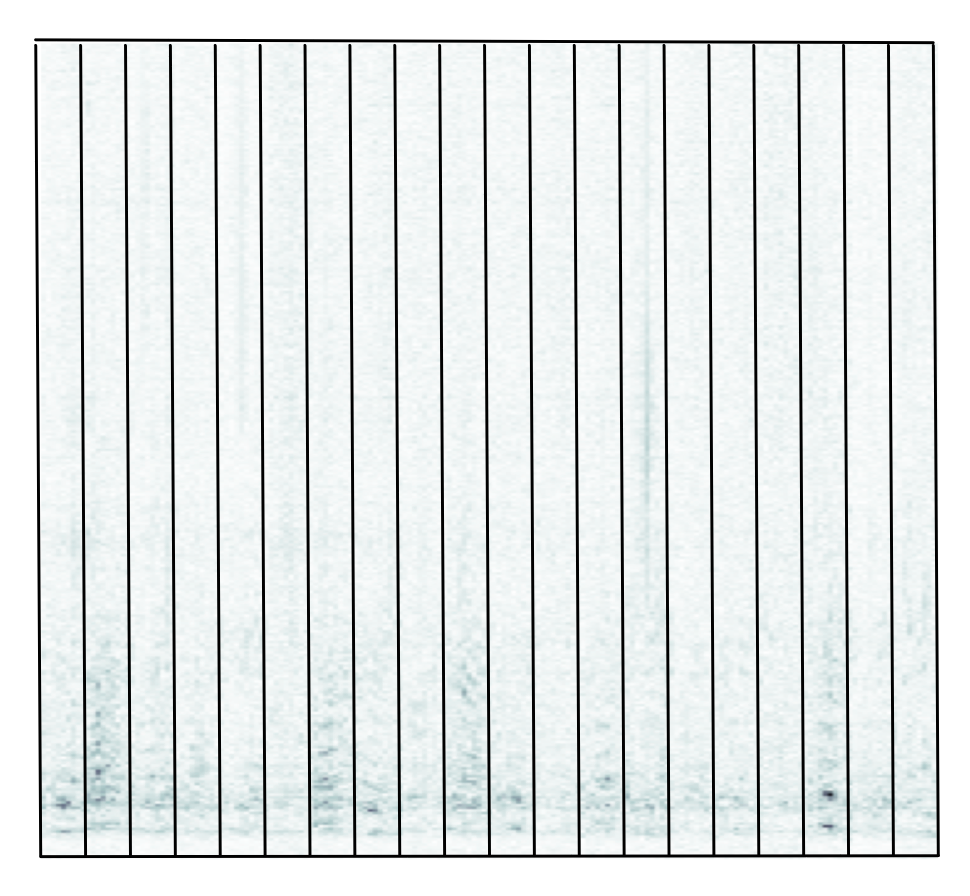}
  \caption{A subset of decoder filters obtained by training the CAE on magnitude-spectrograms of utterances of a male speaker. This decomposition is obtained for the configuration $r = 80$ and $T = 8$. We see that the filters resemble snippets of a speech spectrogram.}~\label{fig:cnn_demo_speech}
\end{figure}
Although the filters can assume negative values, the use of a non-linearity does not allow cross-cancellations across the basis elements.

\subsection{Practical Considerations}
\label{subsec:practical}
Having developed the CCAE equivalent to conv-NMF, we can now begin to understand the nature of the bases and activations learned by the network. To do so, we train the CCAE defined by~(\ref{eq:nmfcnncnn}) on a simple toy example as shown in figure~\ref{fig:cnn_demo}. The input is a spectrogram-like image that consists of a repeating pattern of diagonal structures and has a size of $40 \times 350$ pixels. We use this spectrogram to train a CCAE with filters of size $40 \times 36$. We also incorporate sparsity constraints on the activation, i.e., the output of the first CNN. As shown in the figure, the basis of the decomposition learned by the decoder CNN resembles a snippet of the input spectrogram. As expected, the decoder filters take negative values unlike conv-NMF bases. We also see that the activation comprises a series of impulse trains. Thus, the encoder acts as a matched-filter and identifies the points in time when the corresponding pattern becomes active. As shown in (\ref{eq:nmfcnncnn}), the time-frequency pattern is captured by the filters of the decoder. Given the nature of the activation, we see that the encoder attempts to learn the inverse filter to the decoder. Figure~\ref{fig:cnn_demo_speech} shows the decoder filters obtained by training the CAE on speech utterances of a male speaker. Similar to the previous toy-example, the decoder filters learn patterns that resemble snippets of a speech spectrogram.

Note that the encoder attempts to approximate the inverse of the decoder. From our knowledge of linear filtering in signal processing, we know that the inverse of a finite length filter is given by a recursive filter\footnote{\url{https://ccrma.stanford.edu/~jos/fp/Inverse\_Filters.html}}~\cite{smith2011spectral}. Following this analogy, in the next section we explore using a recurrent neural network in the encoder.

\section{Using a recurrent filter in the encoder}
\label{sec:rnncnn}
In this section, the goal is to construct a recurrent encoder analogous to the convolutional encoder we discussed in the previous section. We will refer to this auto-encoder as a Recurrent-Convolutional Auto-encoder~(RCAE). The potential gain of using a recurrent encoder over a finite length convolutional encoder is due to the fact that a recurrent filter in theory can capture arbitrarily long temporal dependencies. From a signal processing point of view the motivation for going with a recurrent filter is that, the inverse of an finite length filter (the convolutive basis in the encoder) is given by a recurrent filter.

\begin{figure}[t]
  \centering
  \begin{tikzpicture}[auto, node distance=2cm,>=latex']
      
      \node [,name=input] (input) {$\mathbf X$};
      \coordinate [, name=rinput, right of=input, node distance=1.3cm ](rinput) {};
      \node [block, right of=rinput, node distance=1.3cm] (controller) {$\vdots$};
      \node [sum, right of=controller, node distance=2cm] (controllers){$\sum_j$};
      \node [block, above of=controller,node distance=1.3cm] (up){$\text{RNN}_1(\mathbf X)$};
      \node [sum, right of=up, node distance=2cm] (ups) {$\sum_j$};
      \node [block, below of=controller,node distance=1.3cm] (rate) {$\text{RNN}_K(\mathbf X)$};
      \node [sum, right of=rate, node distance=2cm] (rates) {$\sum_j$};	

      \node [sum, right of=controllers,node distance=1.3cm] (sum2) {$g(.)$};
      \node[, right of=sum2,node distance=1.3cm] (system) {$\mathbf H$};
      
      \draw [-] (input) -- (rinput);
      \draw [->] (rinput) -- (controller); 
      \draw [->] (controller) -- (controllers);
      \draw [->] (controllers) -- (sum2);
      \draw [->] (sum2) --(system);
      \draw [->] (rinput) |- (rate);
      \draw [->] (rate) -- (rates);
      \draw [->] (rates) -| (sum2);
      \draw [->] (rinput) |- (up);
      \draw [->] (up) -- (ups);
      \draw [->] (ups) -| (sum2);
      \end{tikzpicture}
  \caption{Block Diagram of RNN Encoder} 
  \label{diag:rnnenc}
\end{figure}
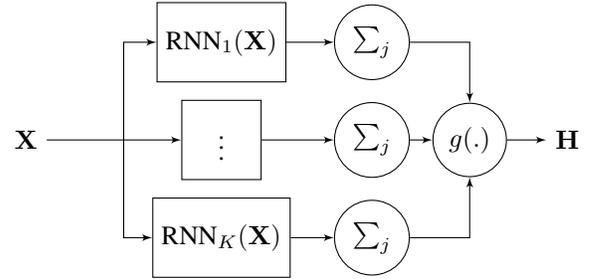

The way we go about building an encoder is by passing the input through $K$ separate recurrent neural networks (RNNs) (Note that $K$ was the number of filters/components in the convolutive model). The $k$'th RNN recursion in the encoder is given by the following equation:    

\begin{align}
  \mathbf Z(k_1, t, k) =& \tanh \Bigg( \sum_{k_2 = 1}^{K_{in}} \mathbf{W^{\ddagger}}^k(k_1,k_2 ) \mathbf{Z}(k_2, t-1, k) + \notag \\
  & \hspace{0cm} \sum_l \mathbf{U^{\ddagger}}^k(k_1,l) \mathbf X(l, t) \Bigg),\; k \in \{1,\dots,K\}, 
\end{align}
where $\mathbf Z(:, t, k) \in \mathbb R^{K_{in}}$ denotes the latent state vector of the $k$'th RNN at time $t$. We denote the hidden state dimensionality of each RNN with $K_{in}$. The recurrent and projection matrices of the $k^{\text{th}}$ RNN are respectively denoted with $\mathbf{W^{\ddagger}}^k$, and $\mathbf{U^{\ddagger}}^k$. Note that although the given recursion corresponds to the vanilla-RNN architecture, there is no restriction on the RNN architecture choice. In our experiments, we have used the LSTM architecture \cite{Hochreiter1997Long,Graves2005Framewise}. After going through the RNN recursions, the encoder output $\mathbf H(i, t)$ is obtained by summing the RNN outputs over the first dimension:
\begin{align}
  \mathbf H(i, t) = \sum_{k_1=1}^{K_{in}} \mathbf Z(k_1, t, i)
\end{align}
The recurrent encoder's block diagram is given in Figure \ref{diag:rnnenc}.

\section{Supervised Source Separation}
\label{sec:ss}
The problem of supervised source separation is solved as a two-step procedure~\cite{smaragdis2007supervised}. The first step of the procedure is to learn suitable models for a given source. We refer to this step as the training step. In the second step, we use these models to explain the contribution of the source in an unknown mixture. In sections~\ref{sec:conv-nmf} and~\ref{sec:rnncnn}, we have developed the auto-encoder architecture to learn suitable convolutive models for a given source. We now turn our attention to the problem of using the models for separating the source in an unknown mixture.

The previous approach to source separation using feed-forward neural-networks~\cite{smaragdis2017aneural} involves estimating the latent representation of the bases. The latent representation captures the contribution of the bases to each frame of the mixture spectrogram. In this approach, we do not utilize the encoder of the trained auto-encoder network for the separation task. In this paper, we present a novel-separation scheme that utilizes the complete auto-encoder architecture for separation. We do so by using the following setup for separation. Given an input spectrogram $\mathbf{X}$, the auto-encoder produces an approximation of the input spectrogram which is a linear combination of its weights. We will denote to this approximation as,
\begin{equation}
    \hat{\mathbf{X}} = Ae(\mathbf{X}|\theta)
    \label{eq:separation_ae}
\end{equation}
Here, $\theta$ denotes the weights (parameters) of the auto-encoder. For the separation procedure, given the trained auto-encoders, (i.e., given $\theta_{1}$ and $\theta_{2}$), the goal is to identify suitable input spectrograms $\mathbf{X}_{1}$ and $\mathbf{X}_{2}$ such that,
\begin{equation}
    \mathbf{X}_{m} = Ae(\mathbf{X}_{1}|\theta_{1}) + Ae(\mathbf{X}_{2}|\theta_{2}) 
    \label{eq:separation}
\end{equation}
In this equation, $\mathbf{X}_{m}$ represents the spectrogram of the mixture and $\mathbf{X}_{1}$,~$\mathbf{X}_{2}$ denote the separated source spectrograms. Thus, similar to NMF, this approach assumes that the magnitude-spectrogram of the mixture is the sum of magnitude-spectrograms of the underlying sources. However, in this separation procedure, we directly estimate the source magnitude-spectrograms without estimating the latent representation. To do so, we train the network defined by~(\ref{eq:separation}) for an appropriate input $\mathbf{X}_{1}$,~$\mathbf{X}_{2}$, instead of training for the weights of the network. As before, we minimize the KL divergence between mixture spectrogram $\mathbf{X}_{m}$ and its approximation~$(\mathbf{X}_{1} + \mathbf{X}_{2})$. Conceptually, this problem is not different to training a neural network. The equivalence can be seen by applying a transposition to the auto-encoder definitions in (\ref{eq:nmfcnncnn}). This approach provides a generalized separation procedure that can be used even when the underlying auto-encoder architectures are changed.

Having obtained the contributions of the sources (separated spectrograms), the next step is to transform these spectrograms back into the time domain. This is given as,
\begin{equation}
    x_{i}(t) = \text{STFT}^{-1}\left(\frac{\mathbf{X_{i}}}{\sum_{i}\mathbf{X}_{i}}\odot \mathbf{X}_{m} \odot e^{i\Phi_{m}}\right)~\text{for}~i\in\{1, 2\}
\end{equation}
Here $x_{i}(t)$ denotes the separated speech signal in time and $\Phi_{m}$ represents the phase of the mixture and $\text{STFT}^{-1}$ is the inverse short-time Fourier transform operation that transforms the complex spectrogram into its corresponding time domain representation. Also, $\odot$ represents the element-wise multiplication operation and the division is also element-wise.

\begin{figure*}[ht!]
\centering
  \includegraphics[clip, trim = 0cm 0cm 0cm 0cm, width=\linewidth]{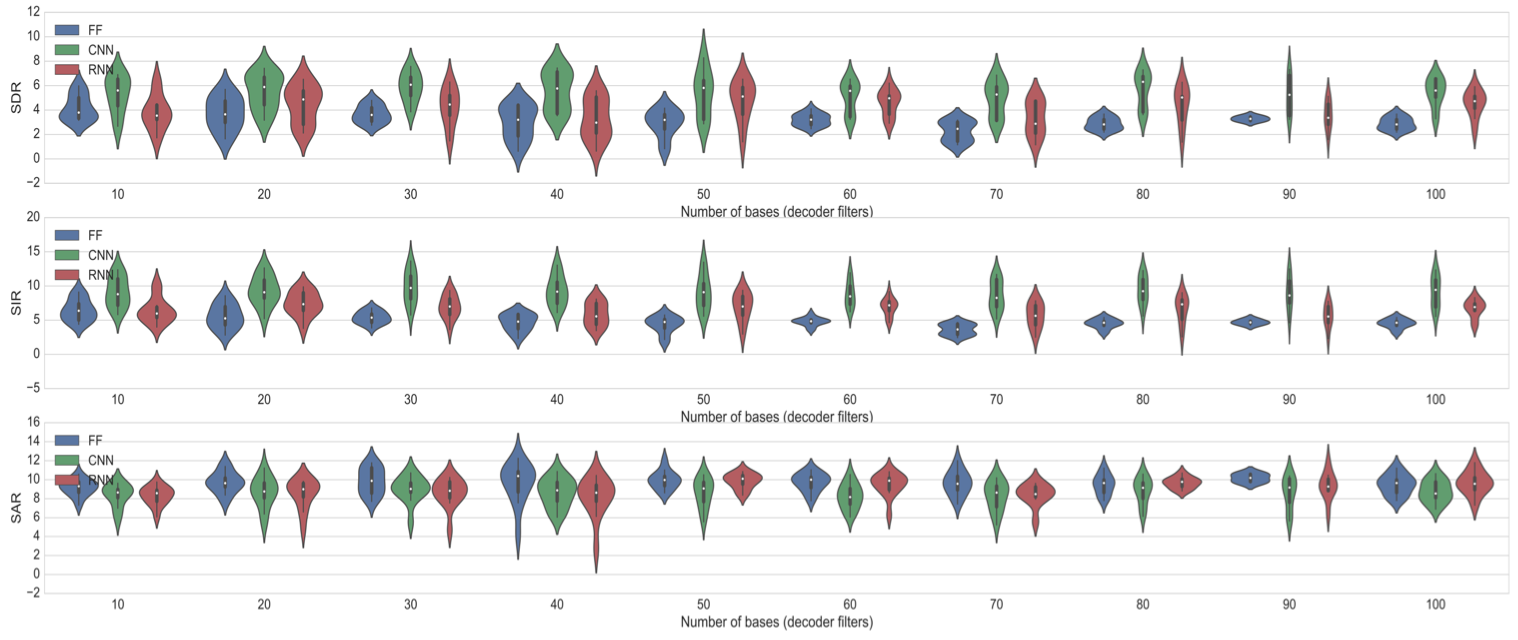}
  \caption{Separation performance of convolutive models obtained using CAE~(centre) and RCAE~(right) for varying numbers of bases ($K$). We compare these models to feed-forward auto-encoder based models~(left) in~\cite{smaragdis2017aneural}. The legend indicates the encoder of the corresponding architectures. }~\label{fig:cnnseparation_performance}
\end{figure*}

\section{Experiments}
\label{sec:experiments}
We now describe the experimental setup used to evaluate our auto-encoder based convolutive audio models. We construct a set of training and test examples using the TIMIT corpus~\cite{timit} for the evaluation. To form the examples, we randomly select a pair of male-female speakers from the TIMIT corpus. Of the $10$ utterances available for each speaker, one utterance is randomly selected for each speaker. These two selected utterances are mixed at $0~dB$ to generate the testing mixture. The remaining $9$ utterances are used as training data to construct models for the sources. In other words, these examples are used to train the auto-encoders. For the evaluation, we generate $20$ such mixtures and compare the models for different parameter configurations. As a pre-processing step, we apply a $1024$ point short-time Fourier transform representation with a hop of $25\%$. The magnitude spectrogram is then given as an input to the network.

The networks are trained by applying a batch gradient descent training procedure and the parameters updated using the RMSProp algorithm~\cite{tieleman2012rmsprop}, with a learning rate and momentum of $0.001$ and $0.7$ respectively. The neural networks are initialized using the Xavier initialization scheme~\cite{glorot2010understanding}.

The CNN filters are selected to be $512$-point tall and $8$-point wide. Thus, the convolutions are performed only along the time axis. The number of CNN filters also decides the number of components in the decomposition. We evaluate these models over a varying number of CNN filters ranging from $10$ to $100$ uniformly in steps of $10$. We compare the separation performance in terms of median BSS\_eval metrics~\cite{fevotte2005bss_eval} viz., signal-to-distortion (SDR), signal-to-interference (SIR) and signal-to-artifact ratio (SAR) parameters. The code for our experiments can be downloaded from \url{https://github.com/ycemsubakan/sourceseparation_nn}. 

\subsection{Results and Discussion}
\label{subsec:results}
Figure~\ref{fig:cnnseparation_performance} gives the separation performance for the CCAE models (centre) and RCAE models (right) for varying values of number of filters~$K$. We evaluate the proposed models by comparing the separation performance to their equivalent feed-forward~(FF) counterparts (left) proposed in~\cite{smaragdis2017aneural}. In order to maintain a uniform experimental setup, we apply the separation scheme described in~\ref{sec:ss} to all the models. We plot the results in terms of a violin plot. The white dots at the centre denote the median value and the thick line denotes the inter-quartile range for each $K$.

We see that the CCAE models significantly out-perform their corresponding FF versions. This can be seen from the fact that the inter-quartile range in SDR for CAE models is higher than the inter-quartile range of corresponding FF models for several values of $K$. This improvement is a consequence of reduced interference generated by these models in source separation (as seen in the SIR plots). Although the SAR for CCAE models degrades slightly as compared to FF models, the significant improvement in SIR compensates for this loss. The convolutive speech models obtained by training the RCAE also outperform the FF auto-encoder models significantly, as seen by the median values and inter-quartile ranges. This improvement in performance is not as pronounced as the CCAE models for the case of speech mixtures. However, the experimentation serves as a definite proof of concept that exploring non-uniform auto-encoder architectures could lead to interesting and potentially powerful algorithms for other datasets and applications. 

The performance of the convolutive models achieves a peak value for $K=80$. However, the median separation performance does not degrade significantly for other values of $K$. Thus, the choice of $K$ does not appear to be a very critical consideration for auto-encoder based convolutive models unlike FF models. At the same time, the variance in SDR seems to be dependent on $K$. We observe that the variance in SDR is considerably lower for higher values of $K$ ($K\geq50$) as seen by the separation results for both the convolutive models. For $K\geq80$, we note that the variance continues to be relatively small even though the spread of the violin plot is large, as shown by the inter-quartile ranges. This implies that the violin plots spread out due to the effect of a few outlier values. In other words, auto-encoder based convolutive audio models produce superior results consistently for a high number of CNN filters.

\vspace{-2mm}
\section{Conclusions}
\label{sec:conclusion}
In this paper, we developed and investigated the use of a convolutional auto-encoder as an alternative to learn convolutive basis decompositions of speech and audio signals. The ability of the networks to include temporal dependencies allow the auto-encoders to learn cross-frame structures in the input spectrogram. We demonstrated that this results in a significant improvement in separation performance as compared to feed-forward auto-encoder models. This approach also allows for several extensions and generalizations to convolutive audio models, that can be easily implemented using the modeling flexibility that comes with neural networks. One such extension considered in this paper is the use of auto-encoders formed by a cascade of recurrent and convolutional layers. Although these models have not outperformed the CAE models for separation of speech mixtures, we have shown that these models are significantly superior to feed-forward auto-encoder models.

\bibliographystyle{IEEEbib}
\bibliography{refs.bib}
\end{document}